\documentclass[conference]{IEEEtran}
\IEEEoverridecommandlockouts
\usepackage{cite}
\usepackage{amsmath,amssymb,amsfonts}
\usepackage{algorithmic}
\usepackage{graphicx}
\usepackage{subcaption}
\usepackage{booktabs}
\usepackage{textcomp}
\usepackage{xcolor}
\usepackage{array}
\usepackage{graphicx}
\usepackage{textcomp}
\usepackage{xcolor}
\usepackage[utf8]{inputenc}
\usepackage[justification=centering, font=footnotesize]{caption}
\usepackage[acronyms,nonumberlist,nopostdot,nomain,nogroupskip]{glossaries}

\newacronym{3gpp}{3GPP}{3rd Generation Partnership Project}
\newacronym{adc}{ADC}{Analog to Digital Converter}
\newacronym{5g}{5G}{5th generation}
\newacronym{aimd}{AIMD}{Additive Increase Multiplicative Decrease}
\newacronym{am}{AM}{Acknowledged Mode}
\newacronym{amc}{AMC}{Adaptive Modulation and Coding}
\newacronym{aqm}{AQM}{Active Queue Management}
\newacronym{awgn}{AWGN}{Additive White Gaussian Noise}
\newacronym{balia}{BALIA}{Balanced Link Adaptation}
\newacronym{bdp}{BDP}{Bandwidth-Delay Product}
\newacronym{qos}{QoS}{Quality of Service}
\newacronym{pqos}{PQoS}{Predictive Quality of Service}
\newacronym{aqosa}{AQoSA}{Agile QoS Adaptation}
\newacronym{bf}{BF}{Beamforming}
\newacronym{cc}{CC}{Congestion Control}
\newacronym{cdf}{CDF}{Cumulative Distribution Function}
\newacronym{cn}{CN}{Core Network}
\newacronym{cqi}{CQI}{Channel Quality Information}
\newacronym{cp}{CP}{Control Plane}
\newacronym{csirs}{CSI-RS}{Channel State Information - Reference Signal}
\newacronym{dc}{DC}{Dual Connectivity}
\newacronym{dce}{DCE}{Direct Code Execution}
\newacronym{dl}{DL}{Downlink}
\newacronym{dmr}{DMR}{Deadline Miss Ratio}
\newacronym{dmrs}{DMRS}{DeModulation Reference Signal}
\newacronym{e2e}{E2E}{end-to-end}
\newacronym{ecn}{ECN}{Explicit Congestion Notification}
\newacronym{edf}{EDF}{Earliest Deadline First}
\newacronym{enb}{eNB}{evolved Node Base}
\newacronym{epc}{EPC}{Evolved Packet Core}
\newacronym{es}{ES}{Edge Server}
\newacronym{fdma}{FDMA}{Frequency Division Multiple Access}
\newacronym{fdd}{FDD}{Frequency Division Duplexing}
\newacronym[firstplural=Radio Access Technologies (RATs)]{rat}{RAT}{Radio Access Technology}
\newacronym{fs}{FS}{Fast Switching}
\newacronym{ftp}{FTP}{File Transfer Protocol}
\newacronym{gnn}{GNN}{Graph Neural Network}
\newacronym{gnb}{gNB}{Next Generation NodeB}
\newacronym{gbr}{GBR}{Guaranteed Bit Rate}
\newacronym{harq}{HARQ}{Hybrid Automatic Repeat request}
\newacronym{hetnet}{HetNet}{Heterogeneous Network}
\newacronym{hh}{HH}{Hard Handover}
\newacronym{hol}{HOL}{Head-of-Line}
\newacronym{ia}{IA}{Initial Access}
\newacronym{ieee}{IEEE}{Institute of Electrical and Electronics Engineers}
\newacronym{imt}{IMT}{International Mobile Telecommunication}
\newacronym{iot}{IoT}{Internet of Things}
\newacronym{ldpc}{LDPC}{Low-Density Parity Check}
\newacronym{los}{LOS}{Line-of-Sight}
\newacronym{lte}{LTE}{Long Term Evolution}
\newacronym{m2m}{M2M}{Machine to Machine}
\newacronym{ml}{ML}{Machine Learning}
\newacronym{mac}{MAC}{Medium Access Control}
\newacronym{mc}{MC}{Multi-Connectivity}
\newacronym{mcs}{MCS}{Modulation and Coding Scheme}
\newacronym{mec}{MEC}{Mobile Edge Cloud}
\newacronym{mi}{MI}{Mutual Information}
\newacronym{mimo}{MIMO}{Multiple Input, Multiple Output}
\newacronym{mmwave}{mmWave}{millimeter wave}
\newacronym{mptcp}{MPTCP}{Multipath TCP}
\newacronym{mr}{MR}{Maximum Rate}
\newacronym{mss}{MSS}{Maximum Segment Size}
\newacronym{mtd}{MTD}{Machine-Type Device}
\newacronym{mtu}{MTU}{Maximum Transmission Unit}
\newacronym{nfv}{NFV}{Network Function Virtualization}
\newacronym{nlos}{NLOS}{Non-Line-of-Sight}
\newacronym{nlosv}{NLOSv}{Vehicle Non-Line-of-Sight}
\newacronym{nr}{NR}{New Radio}
\newacronym{ofdm}{OFDM}{Orthogonal Frequency Division Multiplexing}
\newacronym{pdcch}{PDCCH}{Physical Downlink Control Channel}
\newacronym{sps}{SPS}{semi-persistent scheduler}
\newacronym{pdcp}{PDCP}{Packet Data Convergence Protocol}
\newacronym{pdsch}{PDSCH}{Physical Downlink Shared Channel}
\newacronym{pdu}{PDU}{Packet Data Unit}
\newacronym{pf}{PF}{Proportional Fair}
\newacronym{ff}{FF}{Fairness First}
\newacronym{gi}{GI}{guard interval}
\newacronym{pdf}{PDF}{Probability Density Function}
\newacronym{pgw}{PGW}{Packet Gateway}
\newacronym{phy}{PHY}{Physical}
\newacronym{pbch}{PBCH}{Physical Broadcast Channel}
\newacronym[plural=\gls{mme}s,firstplural=Mobility Management Entities (MMEs)]{mme}{MME}{Mobility Management Entity}
\newacronym{prb}{PRB}{Physical Resource Block}
\newacronym{pss}{PSS}{Primary Synchronization Signal}
\newacronym{pscch}{PSCCH}{Physical Sidelink Control Channel}
\newacronym{pucch}{PUCCH}{Physical Uplink Control Channel}
\newacronym{pusch}{PUSCH}{Physical Uplink Shared Channel}
\newacronym{rach}{RACH}{Random Access Channel}
\newacronym{ran}{RAN}{Radio Access Network}
\newacronym{red}{RED}{Random Early Detection}
\newacronym{rf}{RF}{radio frequency}
\newacronym{rlc}{RLC}{Radio Link Control}
\newacronym{rlf}{RLF}{Radio Link Failure}
\newacronym{rrc}{RRC}{Radio Resource Control}
\newacronym{rrm}{RRM}{Radio Resource Management}
\newacronym{rr}{RR}{Round Robin}
\newacronym{rs}{RS}{Remote Server}
\newacronym{rsrp}{RSRP}{Reference Signal Received Power}
\newacronym{rss}{RSS}{Received Signal Strength}
\newacronym{rtt}{RTT}{Round Trip Time}
\newacronym{rw}{RW}{Receive Window}
\newacronym{rx}{RX}{Receiver}
\newacronym{sa}{SA}{standalone}
\newacronym{sack}{SACK}{Selective Acknowledgment}
\newacronym{sap}{SAP}{Service Access Point}
\newacronym{sc}{SC}{Single Carrier}
\newacronym{sch}{SCH}{Secondary Cell Handover}
\newacronym{scoot}{SCOOT}{Split Cycle Offset Optimization Technique}
\newacronym{sdf}{SDF}{Service Data Flow}
\newacronym{sdma}{SDMA}{Spatial Division Multiple Access}
\newacronym{sinr}{SINR}{Signal to Interference plus Noise Ratio}
\newacronym{sl}{SL}{Sidelink}
\newacronym{sm}{SM}{Saturation Mode}
\newacronym{snr}{SNR}{Signal-to-Noise-Ratio}
\newacronym{son}{SON}{Self-Organizing Network}
\newacronym{ss}{SS}{Synchronization Signal}
\newacronym{srs}{SRS}{Sounding Reference Signal}
\newacronym{sss}{SSS}{Secondary Synchronization Signal}
\newacronym{tb}{TB}{Transport Block}
\newacronym{tcp}{TCP}{Transmission Control Protocol}
\newacronym{tdd}{TDD}{Time Division Duplexing}
\newacronym{tdma}{TDMA}{Time Division Multiple Access}
\newacronym{tfl}{TfL}{Transport for London}
\newacronym{tm}{TM}{Transparent Mode}
\newacronym{trp}{TRP}{Transmitter Receiver Pair}
\newacronym{tti}{TTI}{Transmission Time Interval}
\newacronym{ttt}{TTT}{Time-to-Trigger}
\newacronym{tx}{TX}{Transmitter}
\newacronym{ue}{UE}{User Equipment}
\newacronym{ul}{UL}{uplink}
\newacronym{uml}{UML}{Unified Modeling Language}
\newacronym{um}{UM}{Unacknowledged Mode}
\newacronym{utc}{UTC}{Urban Traffic Control}
\newacronym{vm}{VM}{Virtual Machine}
\newacronym{rsrq}{RSRQ}{Reference Signal Received Quality}
\newacronym{rssi}{RSSI}{Received Signal Strength Indicator}
\newacronym{crs}{CRS}{Cell Reference Signal}
\newacronym{nsa}{NSA}{Non Stand Alone}
\newacronym{mrdc}{MR-DC}{Multi \gls{rat} \gls{dc}}
\newacronym{endc}{EN-DC}{E-UTRAN-\gls{nr} \gls{dc}}
\newacronym{5gc}{5GC}{5G Core}
\newacronym{si}{SI}{Study Item}
\newacronym{iab}{IAB}{Integrated Access and Backhaul}
\newacronym{wf}{WF}{Wired-first}
\newacronym{hqf}{HQF}{Highest-quality-first}
\newacronym{pa}{PA}{Position-aware}
\newacronym{mlr}{MLR}{Maximum-local-rate}
\newacronym{wbf}{WBF}{Wired Bias Function}
\newacronym{mib}{MIB}{Master Information Block}
\newacronym{sib}{SIB}{Secondary Information Block}
\newacronym{rnti}{RNTI}{Radio Network Temporary Identifier}
\newacronym{dft}{DFT}{Discrete Fourier Transform}
\newacronym{kpi}{KPI}{Key Performance Indicator}
\newacronym{ppp}{PPP}{Poisson Point Process}
\newacronym{v2v}{V2V}{Vehicle-to-Vehicle}
\newacronym{wave}{WAVE}{Wireless Access in Vehicular Environments}
\newacronym{udp}{UDP}{User Datagram Protocol}
\newacronym{upa}{UPA}{Uniform Planar Array}
\newacronym{fec}{FEC}{Forward Error Correction}
\newacronym{v2x}{V2X}{Vehicle-To-Everything}
\newacronym{psfch}{PSFCH}{Physical Sidelink Feedback Channel}
\newacronym{pssch}{PSSCH}{Physical Sidelink Shared Channel}
\newacronym{csma}{CSMA}{Carrier Sense Multiple Access}
\newacronym{v2n}{V2N}{Vehicle-to-Network}
\newacronym{wlan}{WLAN}{Wireless Local Area Network}
\newacronym{cav}{CAV}{Connected and Autonomous Vehicle}
\newacronym{v2i}{V2I}{Vehicle-to-Infrastructure}
\newacronym{d2d}{D2D}{Device-to-Device}
\newacronym{c-its}{C-ITS}{Connected Intelligent Transportation System}
\newacronym{fr2}{FR2}{Frequency Range 2}
\newacronym{fr1}{FR1}{Frequency Range 1}
\newacronym{bs}{BS}{Base Station}
\newacronym{sdu}{SDU}{Service Data Unit}
\newacronym{csi}{CSI}{Channel State Information}
\newacronym{scs}{SCS}{Subcarrier Spacing}
\newacronym{sumo}{SUMO}{Simulation of Urban MObility}
\newacronym{prr}{PRR}{Packet Reception Ratio}
\newacronym{edca}{EDCA}{Enhanced Distribution Channel Access}
\newacronym{sdap}{SDAP}{Service Data Adaptation Protocol}
\newacronym{iiot}{IIoT}{Industrial Internet of Things}
\newacronym{agv}{AGV}{Automated Guided Vehicles}
\newacronym{cm}{C/M}{Controller/Master}
\newacronym{soa}{SoA}{State-of-the-Art}
\newacronym{snpn}{SNPN}{Standalone Non-Public Network}
\newacronym{pninpn}{PNI-NPN}{Public Network Interface Non-Public Network}
\newacronym{urllc}{URLLC}{Ultra-Reliable Low-Latency Communications}
\newacronym{qci}{QCI}{QoS Class Identifier}
\newacronym{ai}{AI}{Artificial Intelligence}
\newacronym{mab}{MAB}{Multi-Armed Bandit}
\newacronym{su}{SU}{Scheduling Unit}
\newacronym{ra}{RA}{Random Agent}
\newacronym{na}{NA}{Neural Agent}
\newacronym{ucba}{UCB-A}{UCB Agent}
\newacronym{tsa}{TS-A}{TS Agent}
\newacronym{ucb}{UCB}{Upper Confidence Bound}
\newacronym{ts}{TS}{Thompson Sampling}
\newacronym{nlts}{NLTS}{Neural Linear Thompson Sampling}
\newacronym{inf}{InF}{Indoor Factory}
\newacronym{infsl}{InF-SL}{Indoor Factory - Sparse Clutter, Low BS}
\newacronym{infdl}{InF-DL}{Indoor Factory - Dense Clutter, Low BS}
\newacronym{infsh}{InF-SH}{Indoor Factory - Sparse Clutter, High BS}
\newacronym{infdh}{InF-DH}{Indoor Factory - Dense Clutter, High BS}
\newacronym{us}{US}{Uplink Scheduler}
\newacronym{nn}{NN}{Neural Network}
\newacronym{dnn}{DNN}{Deep Neural Network}
\newacronym{das}{DAS}{Distributed Antenna System}
\newacronym{rb}{RB}{Resource Block}
\newacronym{rl}{RL}{Reinforcement Learning}
\newacronym{embb}{eMBB}{enhanced Mobile BroadBand}
\newacronym{5gacia}{5G-ACIA}{5G Alliance for Connected Industries and Automation}
\newacronym{cnn}{CNN}{Convolutional Neural Network}
\newacronym{aipm}{AIPM}{Artificial Intelligence Program Manager}
\newacronym{fci}{FCI}{Feedback Control Information}
\newacronym{prach}{PRACH}{Physical Random Access Channel}
\newacronym{coreset}{CORESET}{Control Resource Set}
\newacronym{cmab}{CMAB}{Contextual Multi-Armed Bandit}
\newacronym{drl}{DRL}{Deep Reinforcement Learning}
\newacronym{uav}{UAV}{Unmanned Aerial Vehicle}
\newacronym{ssps}{SSPS}{smart semi-persistent scheduler}
\newacronym{asps}{ASPS}{adaptive semi-persistent scheduler}
\newacronym{osps}{OSPS}{original semi-persistent scheduler}
\newacronym{bsps}{BSPS}{baseline semi-persistent scheduler}
\newacronym{fd}{FD}{frequency domain}
\newacronym{td}{TD}{time domain}
\newacronym{mno}{MNO}{Mobile Network Operator}
\newacronym{qoe}{QoE}{Quality of Experience}
\newacronym{bler}{BLER}{Block Error rate}
\newacronym{bl}{BL}{Bayesian Learning}
\newacronym{gml}{GML}{Machine Learning on Graphs}
\newacronym{bnn}{BNN}{Bayesian Neural Network}
\newacronym{pnn}{PNN}{Probabilistic Neural Network}
\newacronym{svi}{SVI}{Stochastic Variational Inference}
\newacronym{mape}{MAPE}{Mean Absolute Percentage Error}
\newacronym{ae}{AE}{Autoencoder}
\newacronym{vae}{VAE}{Variational Autoencoder}
\newacronym{lstm}{LSTM}{Long-Short Term Memory}
\newacronym{rnn}{RNN}{Recurrent Neural Network}
\newacronym{dci}{DCI}{Downlink Control Information}
\newacronym{rrf}{RF}{Random Forest}
\newacronym{mlp}{MLP}{Multi-layer Perceptron}
\def\BibTeX{{\rm B\kern-.05em{\sc i\kern-.025em b}\kern-.08em
    T\kern-.1667em\lower.7ex\hbox{E}\kern-.125emX}}
    \usepackage{amsmath,amssymb,amsfonts}

\begin{document}
\bstctlcite{IEEEexample:BSTcontrol}

\title{Data-driven Predictive Latency for 5G: A Theoretical and Experimental Analysis Using Network Measurements\vspace{-.3cm}}
\author{\IEEEauthorblockN{
Marco Skocaj\IEEEauthorrefmark{1},
Francesca Conserva\IEEEauthorrefmark{1},
Nicol Sarcone Grande\IEEEauthorrefmark{1},
Andrea Orsi\IEEEauthorrefmark{2},
Davide Micheli\IEEEauthorrefmark{2},\\
Giorgio Ghinamo\IEEEauthorrefmark{2},
Simone Bizzarri\IEEEauthorrefmark{2} and
Roberto Verdone\IEEEauthorrefmark{1}}\\
\vspace{-.4cm}
\IEEEauthorblockA{\IEEEauthorrefmark{1}
\small DEI, University of Bologna, \& WiLab, CNIT, \textit{Italy}}
\IEEEauthorblockA{\IEEEauthorrefmark{2}
\small TIM, \textit{Italy}}
\thanks{This work has been accepted for publication at IEEE PIMRC 2023.}\thanks{© 2023 IEEE.  Personal use of this material is permitted.  Permission from IEEE must be obtained for all other uses, in any current or future media, including reprinting/republishing this material for advertising or promotional purposes, creating new collective works, for resale or redistribution to servers or lists, or reuse of any copyrighted component of this work in other works}
\vspace{-1cm}
}

\maketitle

\begin{abstract}
The advent of novel 5G services and applications with binding latency requirements and guaranteed \gls{qos} hastened the need to incorporate autonomous and proactive decision-making in network management procedures. The objective of our study is to provide a thorough analysis of predictive latency within 5G networks by utilizing real-world network data that is accessible to mobile network operators (MNOs). In particular, (i) we present an analytical formulation of the user-plane latency as a Hypoexponential distribution, which is validated by means of a comparative analysis with empirical measurements, and (ii) we conduct experimental results of probabilistic regression, anomaly detection, and predictive forecasting leveraging on emerging domains in \gls{ml}, such as \gls{bl} and \gls{gml}. We test our predictive framework using data gathered from scenarios of vehicular mobility, dense-urban traffic, and social gathering events. Our results provide valuable insights into the efficacy of predictive algorithms in practical applications.
\end{abstract}

\begin{IEEEkeywords}
Predictive Quality of Service, Latency, Machine Learning, Bayesian Learning, Machine Learning on Graphs, 5G. \vspace{-.55cm}
\end{IEEEkeywords}

\section{Introduction}
The \gls{5g} wireless technology allows the virtual connection of everyone and everything together, including machines and devices. \gls{urllc} is one of the leading pillars of the \gls{5g} standard, which aims to provide extremely low latency values and reliability up to 99.99\% \cite{5gaciaiiot}. Industries, transportation, precision agriculture, and \gls{v2x} communications are some of the driving applications for the development of \gls{urllc}. The rise of such new services and applications with binding latency requirements and guaranteed \gls{qos}, together with recent advancements in \gls{ai}, are paving the way to the deployment of autonomous connected systems. Within this context, the idea of \gls{pqos} has been introduced as a means of equipping autonomous systems with proactive notification regarding imminent changes in \gls{qos}.
The experienced \gls{qos} is affected by various elements, such as interference, mobility, network conditions, and terminal characteristics (e.g., number of antennas). Although different services have different \gls{qos} constraints in terms of latency and reliability, being able to prevent a session interruption due to \gls{qos} degradation becomes a key requirement \cite{qos_automated_driving}. In this respect, being able to predict \gls{qos} changes, becomes a crucial aspect to preventively adjust the application behavior \cite{5GAA, predictiveQos_next_frontier}. Nowadays, minimizing latency means providing aid for real-time applications (e.g., online games, autonomous driving, etc.), ensuring greater interactivity and smoother experiences, increasing the energy efficiency of \gls{5g} networks, and improving reliability in mission-critical applications. Furthermore, latency is critical to foster a range of new applications, such as virtual and augmented reality, smart cities, and connected cars. \glspl{mno} have access to a vast amount of \gls{ran} measurements, including network \glspl{kpi} and counters, measuring uplink/downlink data volumes, transmission parameters, monitoring of radio resources, as well as accessibility/handover requests/failures, among many others. Such data availability offers significant opportunities: different levels of granularity at both a spatial and temporal level boost the network analysis capability, enabling the true potential of \gls{pqos}.

The present study aims to provide a thorough investigation of predictive latency within \gls{5g} networks. This is achieved through the utilization of \gls{kpi} \gls{ran} measurements obtained at each \gls{gnb}, combined with the development of a predictive framework based on cutting-edge \gls{ml} methodologies. Our objective is to offer a comprehensive analysis of the key factors affecting predictive latency in 5G networks and to assess the potential benefits of employing advanced \gls{ml} techniques in this context. In this regard, we collected measurements on three clusters characterized by diverse traffic patterns, among which vehicular and dense-urban traffic.


\subsection{State of the art}
Various studies have focused on the subject of \gls{pqos}, with particular emphasis on its relevance to \gls{v2x} communications. Authors in \cite{qos_v2x_ml} model the \gls{qos} prediction as a binary classification problem to determine whether a packet can be delivered within a defined latency window with the use of standard \gls{ml} techniques such as \glspl{rrf} and \glspl{mlp}. In the framework of \gls{aqosa}, a \gls{qos} adjustment assistance mechanism has been developed to predict and notify \gls{qos} changes at application level \cite{qos_automated_driving}. Another aspect of \gls{pqos} concerns the identification of the relationship between features. In this regard, a work of noticeable importance is \cite{routenet}, which proposes a network model based on \glspl{gnn} capable of understanding the connection between topology and input traffic to estimate the per-packet delay distribution using deep learning techniques. With respect to \gls{urllc}, the authors of \cite{qos_fastfading} attempt to monitor and forecast the rapid fluctuations in channel conditions caused by fast fading, in order to facilitate advanced scheduling. Finally, the research community has shown significant interest in reducing latency in 5G networks. Various analytical models have been developed to evaluate \gls{e2e} latency by implementing different scheduling configurations and observing several 5G features \cite{latformula,e2elatency_v2x}.

\subsection{Contributions}
Our work aims to offer a comprehensive analysis of predictive latency in 5G networks using real-world network data available to \glspl{mno} and developing a solid framework leveraging recent advancements in \gls{ml}. Our contributions can be summarized as follows:

\begin{itemize}
    \item Starting from 3GPP definitions, we present an analytical formulation of the U-plane (User-plane) latency, proving the latter can be modeled as a Hypoexponential distribution. We ascertain the validity of our analytical outcomes by means of a comparative analysis with empirical network measurements.
    \item We discuss the use of emerging domains within the field of \gls{ml}, such as \gls{bl} and \gls{gml} to tackle three distinct PQoS use cases: probabilistic regression, anomaly detection, and predictive forecasting.
    \item We conduct numerical experiments using \glspl{kpi} collected from three distinct traffic scenarios, namely vehicular mobility, dense-urban environment, and social gathering events. Our objective is to evaluate the performance of predictive models under diverse and representative traffic conditions.
\end{itemize}

\section{Problem Formulation}
\label{ref_scenario}

\begin{figure}
    \centering
    \includegraphics[width=.75\columnwidth]{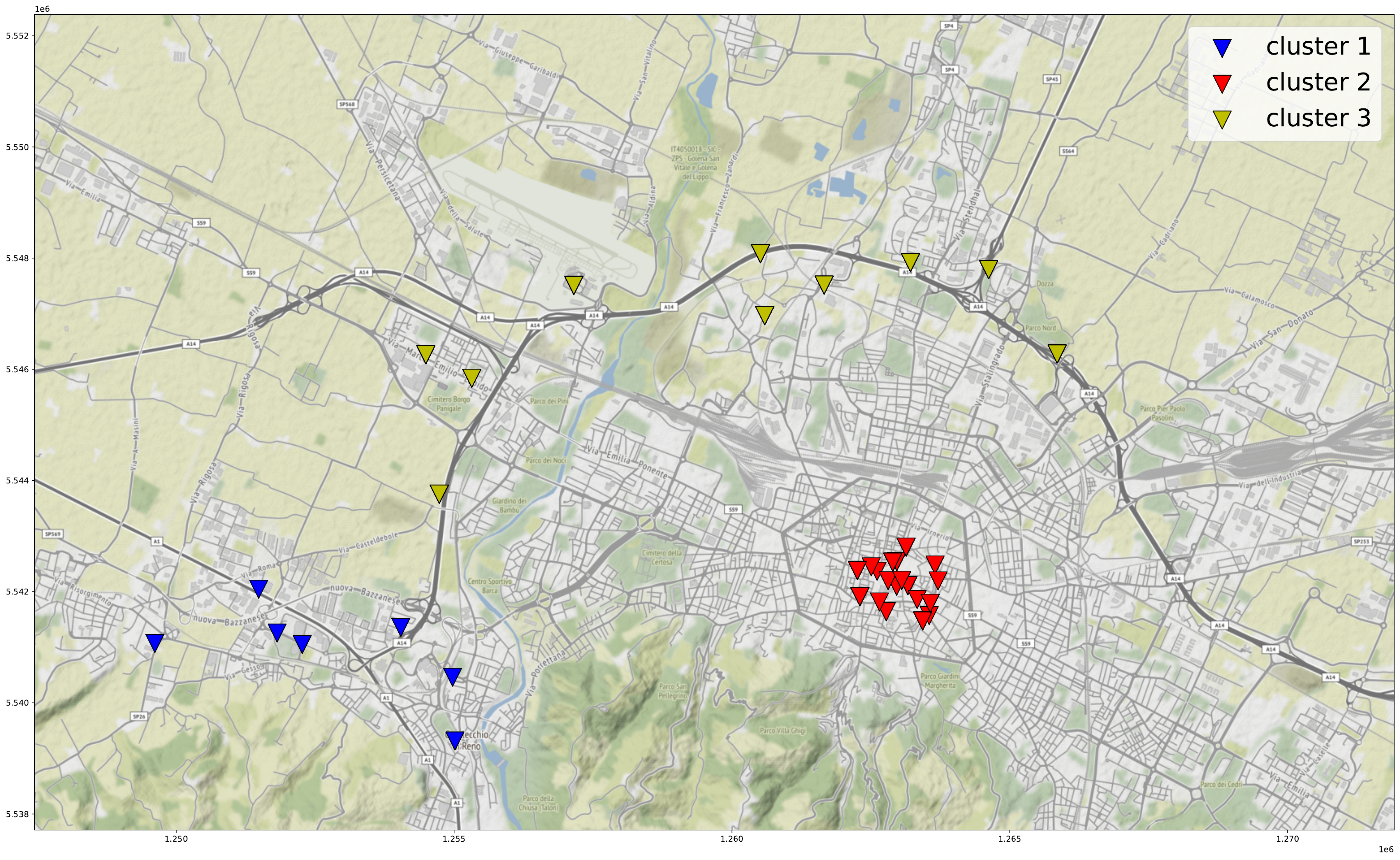}
    \caption{Bologna map, clusters of cells.}
    \label{fig:cell_map}
    \vspace{-.6cm}
\end{figure}

Our reference scenario comprises network \glspl{kpi} gathered from three clusters of cells scattered throughout the entire area of the city of Bologna, Italy (Fig. \ref{fig:cell_map}).
The first cluster gathers \glspl{gnb} from the city center area, the second one encompasses the highway and the ring road, whereas the last one covers an industrial area hosting concerts and social gathering events. The network \glspl{kpi} exploited for the latency prediction are obtained as a statistical average of the measurements gathered with a periodic interval of 15 minutes, for a total of one entire month of data. Details about the feature selection and the individual indicators are discussed in section \ref{latency_kpis}.

\noindent\gls{3gpp} employs the \gls{qci} scalar value to assess the quality of packet communication. The QCI refers to a particular packet forwarding behavior (e.g.: resource type, priority, and packet loss rate) to be delivered to a \gls{sdf} \cite{3gpp.23.203}. For the scope of this work, we focus our attention on \glspl{kpi} data collected from two \gls{qci} classes:
\begin{itemize}
    \item \gls{qci}1, i.e. conversational voice service that requires \gls{gbr} resource type and packet error loss rate of $10^{-2}$.
    \item \gls{qci}7, i.e. voice, video streaming, and interactive gaming services that represent the majority of traffic nowadays; they require non-\gls{gbr} resource type and packet error loss rate of $10^{-3}$.
\end{itemize}

The remainder of this section delves into the details of latency formulation and derives a probabilistic interpretation of the U-plane latency within \gls{5g} communications systems. This model is derived from the definitional framework established by the \gls{3gpp}, described in section \ref{3GPP Overview}.

\subsection{3GPP Overview}
\label{3GPP Overview}
According to \gls{3gpp}, latency can be formalized as the sum of C-plane (Control-plane) and U-plane (User-plane) latency\cite{3gpp.25.912}. The former measures the time elapsed from a \gls{ue}'s \gls{rach} preamble transmission and the successful reception at the \gls{gnb} of a \gls{rrc} Connection Complete message; in other words, it measures the transition time of a \gls{ue} from \gls{rrc}-idle state to \gls{rrc}-connected state. On the other hand, the U-plane latency is a measure of the transit time between a packet being available at the \gls{ue} (or \gls{ran} \gls{gnb}) IP layer, and the availability of this packet at the IP layer of the \gls{ran} \gls{gnb} (or \gls{ue}) \cite{3gpp.25.912}. Besides the processing delays, the \gls{tti} duration, and \gls{harq} loop needed to receive the packet correctly, the U-plane latency also accounts for the number of packet retransmissions occurring with probability equal to the \gls{bler}.

\noindent Within the scope of our work, we direct our attention towards the \gls{dl} U-plane latency for a two-fold reason: (i) the user plane latency offers greater degrees of freedom in terms of optimization compared to the C-plane latency, which depends primarily on the random access procedure; (ii) network measurements are collected uniquely for users in RRC-connected mode, for whom U-plane latency is the sole quantifiable delay because it involves only the \gls{ran}, whereas the C-plane latency includes delay contributions that impact the \gls{cn} as well. Furthermore, the U-plane \gls{dl} represents the majority of generated traffic. As a final remark, we consider the case of dynamic (grant-based) scheduling, for which the \gls{gnb} needs to forward scheduling information to the \gls{ue} before transmitting data on the PDSCH.  




\subsection{Latency formulation}
\label{latency formulation}
\begin{figure}
  \centering
  \begin{subfigure}{0.4\columnwidth}
    \includegraphics[width=\linewidth]{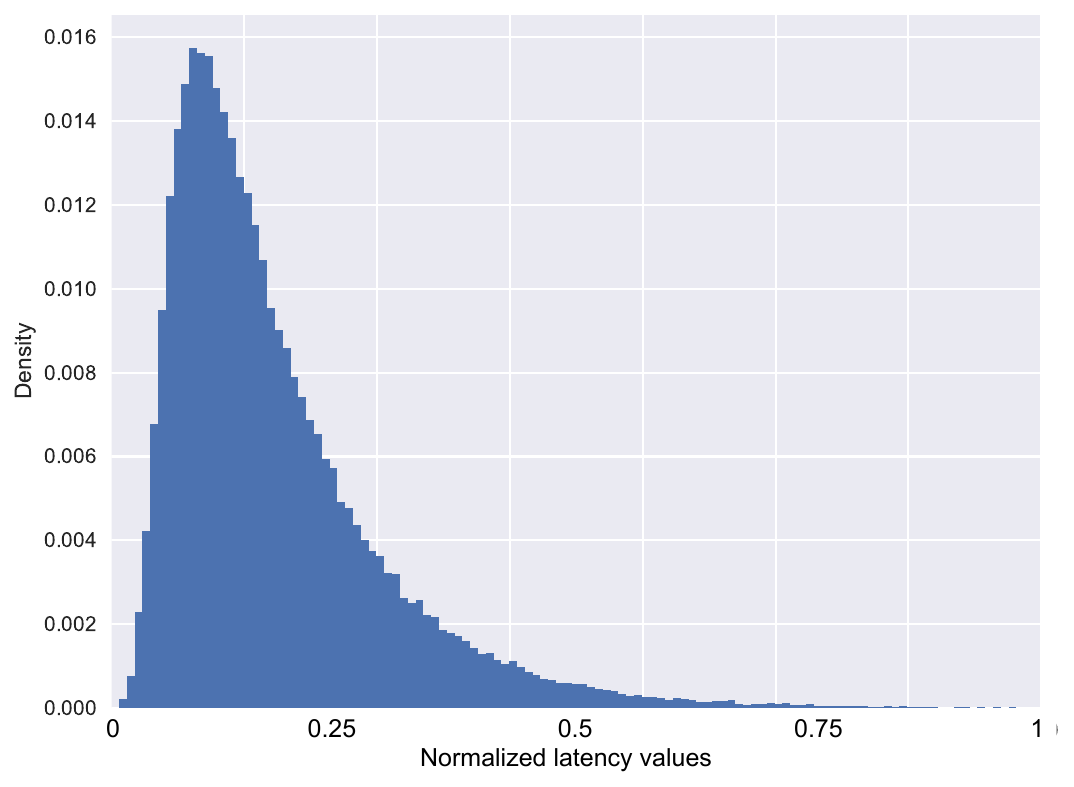}
    \caption{Numerical evaluation of \eqref{eq:P(L2)}.}
    \label{fig:subfig1}
  \end{subfigure}
  \hspace{0.05\textwidth}
  \begin{subfigure}{0.4\columnwidth}
    \includegraphics[width=\linewidth]{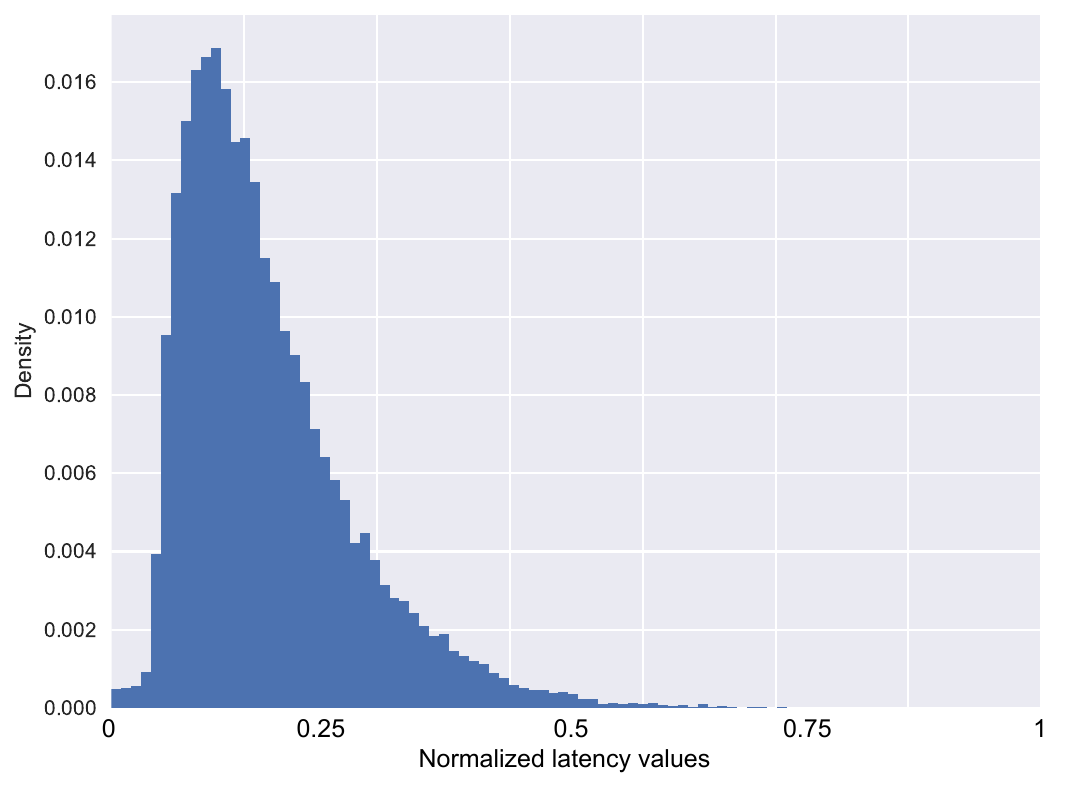}
    \caption{Empirical latency distribution}
    \label{fig:subfig2}
  \end{subfigure}
  \caption{Comparison of empirical and theoretical pdfs: (a) Theoretical formulation, numerically evaluated for $\text{BLER}=0.1$, (b) Empirical latency distribution observed from network \glspl{kpi}.}
  \label{fig:subfigure}
  \label{fig:pdfs}
  \vspace{-.6cm}
\end{figure}

Leveraging on the \gls{3gpp} definitions introduced above, we define the U-plane latency, denoted as $L$, as per \eqref{eq:L2}:
\begin{equation}
\small
    \label{eq:L2}
    L = \tau_{\rm radio} + \tau_{\rm HARQ} + N * (\tau_{\rm radio}' + \tau_{\rm HARQ}),
\end{equation}
with $N=\{0, 1, \dots, N_{\text{max}}\}$. In \eqref{eq:L2}, $\tau_{\rm radio}$ is a random variable accounting for the radio latency over the Uu interface related to the first \gls{gnb}-\gls{ue} transmission. Similarly, $\tau_{\rm radio'}$ accounts for the same delay when a packet is re-scheduled for transmission upon reception of a negative acknowledgment. For the sake of generality, we account for the two terms as separate and independent random variables, assuming that prioritization mechanisms take place for the dynamic scheduling of previously discarded packets, i.e., $\mathbb{E}[\tau_{\rm radio}'] \leq \mathbb{E}[\tau_{\rm radio}]$. Finally,  $\tau_{\rm HARQ}$ accounts for the delay introduced by the HARQ mechanisms, and $N$ denotes the total number of re-transmissions.\\
\noindent Notice that Eq. (\ref{eq:L2}) is a measure of latency at the IP layer, consistent with the \gls{3gpp} definition discussed in Sec. \ref{3GPP Overview}. The present analysis excludes the transport layer due to the tendency to introduce varying additional delays based on the particular protocol employed (e.g., \gls{tcp} introduces delays due to connection establishment between two end-point, or error-checking and re-transmissions of corrupted packets).
\begin{figure}[h]
    \begin{center}
    \includegraphics[width= 0.65\columnwidth,keepaspectratio]{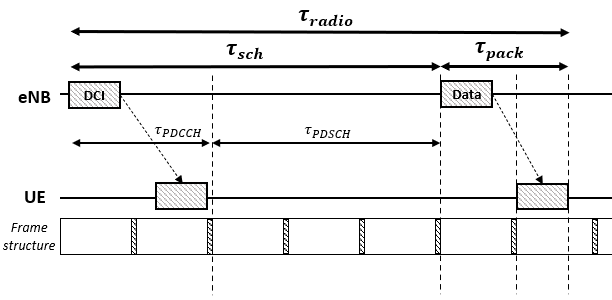}
    \caption{$\tau_{\rm radio}$ for DL transmission.}
    \label{fig:tau_radio}
    \end{center}
    \vspace{-.5cm}
\end{figure}
Similarly to previous works \cite{latformula}, we can hereby decompose $\tau_{\rm radio}~(\tau_{\rm radio}')$ into the sum of two independent quantities related to the scheduling and the transmission time of the packet, namely $\tau_{\rm sch}~(\tau_{\rm sch}')$ and $\tau_{\rm pack}$. Downlink scheduling information on the PDSCH is delivered to the \gls{ue} by the \gls{dci} on the PDCCH. Differently from \cite{latformula}, we denote with $\tau_{\text{sch}}$ the whole interval of time between the packet generation at the \gls{gnb} and the instant when the packet is transmitted on the PDSCH, as reported in Fig.~\ref{fig:tau_radio}. 
Accordingly, $\tau_{\text{sch}}$ coincides with the \textit{waiting time} of a $M/M/1$ system, as per traditional queuing theory. On the other hand, $\tau_{\text{pack}}$, which is fully determined by \gls{5g} numerology, average \gls{ue}'s \gls{mcs}, number of available \glspl{rb}, etc., is equivalent to the \textit{service time}. In Appendix A, leveraging on queuing theory, we show that the sum of $\tau_{\text{sch}}$ and $\tau_{\text{pack}}$ in an $M/M/1$ system can be modeled as a negative exponential distribution. On the other hand, for the sake of simplicity and without loss of generality, let us assume $\tau_{\text{HARQ}}$ as a fixed delay. Consequently, it is possible to re-formulate \eqref{eq:L2} as a sum of independent random variables, as per \eqref{eqn:L2}:
\vspace{0.1pt}
\begin{equation}\label{eqn:L2}
\small
    L = \underbrace{\tau_{\text{sch}} + \tau_{\text{pack}} + \underbrace{\tau_{HARQ}}_{=~C}}_{\tau_{\text{tx}}} + N \cdot (\underbrace{\tau_{\text{sch}}' +  \tau_{\text{pack}} + \underbrace{\tau_{HARQ}}_{=~C}}_{\tau_{\text{rtx}}}),
\end{equation}
where $N$ is geometrically distributed with success parameter $p$ equivalent to the complementary \gls{bler}, i.e., $N \sim geom(1 - \text{BLER})$, and $\tau_{\text{sch}} + \tau_{\text{pack}} \sim exp(\lambda_1)$, $\tau_{\text{sch}}' + \tau_{\text{pack}} \sim exp(\lambda_2)$. As a result, $\tau_{\text{tx}}$ and $\tau_{\text{rtx}}$ are still negative exponential distribution with rate parameters $\lambda_1$, $\lambda_2$ and mean value $\frac{1}{\lambda_i} + C$. Writing $N$ in explicit form, we can reformulate \eqref{eqn:L2} as: 
\begin{equation}\label{eq:P(L2)}
\small
\begin{split}
    L~&=~\sum_{j=0}^{N_{\text{max}}}  P_j(\tau_{\text{tx}} + j \cdot  \tau_{\text{rtx}})=~\tau_{\text{tx}}\underbrace{\sum_{j=0}^{N_{\text{max}}} P_j}_{=1}~+~\tau_{\text{rtx}}\sum_{j=1}^{N_{\text{max}}}j \cdot P_j~= \\&
    =~\tau_{\text{tx}} + \tau_{\text{rtx}}  P_1 + 2\tau_{\text{rtx}}  P_2 + 3\tau_{\text{rtx}}P_3 + \dots + o(P_n),
\end{split}
\end{equation}
where $P_j = P(N=j) = \text{BLER}^j\cdot (1-\text{BLER})$. Thus, $L$ can be approximated at the $n$-th order as the sum between $n$ independent negative exponential random variables with monotonically increasing rate values $\propto 1/P_j$, i.e., $\tau_{\text{tx}} \sim exp(\lambda_1)$ and $\tau_{\text{rtx}} \sim exp(\lambda_2/P_j)$. As analytically shown in Appendix B, this results in a Hypoexponential distribution $L \sim hexp(\lambda_1, \dots, \lambda_N)$. Our theoretical formulation is confirmed by empirical data, as depicted in Fig. \ref{fig:pdfs}.

\section{Measuring Latency from network KPIs}
\label{latency_kpis}


In this section, we will discuss the \glspl{kpi} employed as the ground truth ($L$) in our analysis, as well as the designed feature space. In order to identify a suitable set of \glspl{kpi} for measuring and predicting $L$, it is essential to have a comprehensive understanding of the factors that influence its behavior. As per \eqref{eqn:L2}, $L$ is influenced by both traffic and radio channel conditions. Indeed, situations of high network congestion may adversely affect $\tau_{\text{sch}}$, which is dependent on the number of total \glspl{ue} in the queue. Similarly, unfavorable radio conditions and high levels of interference can lead to an increased \gls{bler}, requiring a potentially higher number of packet retransmissions to achieve a successful transmission. 


\subsection{Ground truth evaluation}
The identified \glspl{kpi} for estimating $L$ provides a measure of the delay in transmitting a \gls{pdcp} \gls{sdu} in the downlink given a specific \gls{qci} value, as defined in the \gls{3gpp} TS 36.314, whose definition is reported in \eqref{eq:packet_delay_3gpp}:
\vspace{0.5pt}
\begin{equation}
\small
\label{eq:packet_delay_3gpp}
    P_{\text{delay}}(T, QCI) = {\left\lfloor \frac{\sum_{i}t_{\text{ack}}(i)-t_{\text{arriv}}(i)}{I(T)} \right\rfloor},
\end{equation}
where $t_{\text{arriv}}(i)$ is the point in time when the \gls{pdcp} \gls{sdu} reaches the \gls{pdcp} layer at the transmitter side (i.e., at the \gls{gnb} in case of DL transmission); $t_{\text{ack}}(i)$ represents the instant corresponding to the last piece of the $i$-th \gls{pdcp} \gls{sdu} received by the \gls{gnb} according to received \gls{harq} feedback. Finally, $I(T)$ indicates the total number of \gls{pdcp} \glspl{sdu}, and $T$ represents the period during which the measurement is performed. 

The elected \gls{kpi} is provided as the average sum of two network counters: the first accounts for the retention delay within the \gls{gnb}; the second considers the average delay introduced by \gls{harq} loop. Although this measure is taken at the \gls{pdcp} level, thus excluding the IP layer latency, without loss of generality, it can still be considered a good estimation of $L$. Indeed, the missing inter-layer processing delay can be neglected if compared to the other delay contributions.

\subsection{Feature selection}
Here, we delve into the feature selection process, where we identify the \glspl{kpi} to be utilized for predicting $L$. Feature selection has been performed following both numerical investigations, such as correlation analysis, and logical criteria. Specifically, we considered the dependency of $L$ on traffic conditions and network quality. The first group includes the average number of active users in the \gls{dl}, the traffic volume (expressed in terms of \gls{pdcp} \glspl{sdu}) in \gls{dl}, and the average \gls{prb} usage during the \gls{tti} in the \gls{dl}. On the other hand, the second group leverages the average \gls{cqi}, the average values of \gls{rssi} and \gls{sinr} on the on \gls{pusch}, and the average values of \gls{mcs} on both the \gls{pusch} and \gls{pdsch}. As an additional feature, the temporal information of data acquisition is also incorporated. Table \ref{tab1} displays the results of a Pearson correlation analysis between the selected features and the latency measure intended for prediction.
It is noteworthy that the features related to traffic exhibit a stronger correlation in comparison to those pertaining to the quality of the radio channel. As a matter of fact, the \glspl{kpi} related to the utilization of resources in the \gls{dl} shows a correlation value of approximately 0.8 with the average \gls{pdcp} \gls{sdu} latency in \gls{dl}.

\begin{table}[h]
\caption{Correlation Analysis}
\setlength{\tabcolsep}{3pt}
\begin{tabular}{|p{120pt}|p{120pt}|}
\hline
\centering
Feature space& 
Pearson's correlation value with the average \gls{pdcp} \gls{sdu} latency in \gls{dl}\\
\hline
Time& 
0.39\\
Traffic volume in DL& 
0.62\\
Resources' utilization in DL per TTI&
0.79\\
Number of active UEs in DL&
0.67\\
Average CQI &
-0.35\\
Average RSSI in UL &
-0.33\\
Average SINR in UL &
-0.33\\
Average MCS in DL &
0.11\\
Average MCS in UL &
-0.47\\
\hline
\end{tabular}
\label{tab1}
\vspace{-.6cm}
\end{table}


\section{Algorithms and Experimental results}
\label{experimental_results}
This section presents exemplary experimental results for three use cases of interest in the context of \gls{pqos}. The subsequent subsections introduce each use case, elucidate the underlying theoretical aspects of the proposed algorithms, and subsequently present the numerical outcomes. To ensure the preservation of sensitive information of the \gls{mno}, the numerical findings are displayed in a standardized format.

\subsection{Use case 1: Bayesian probabilistic regression}
\begin{figure}[h]
    \centering
    \includegraphics[width=.6\columnwidth]{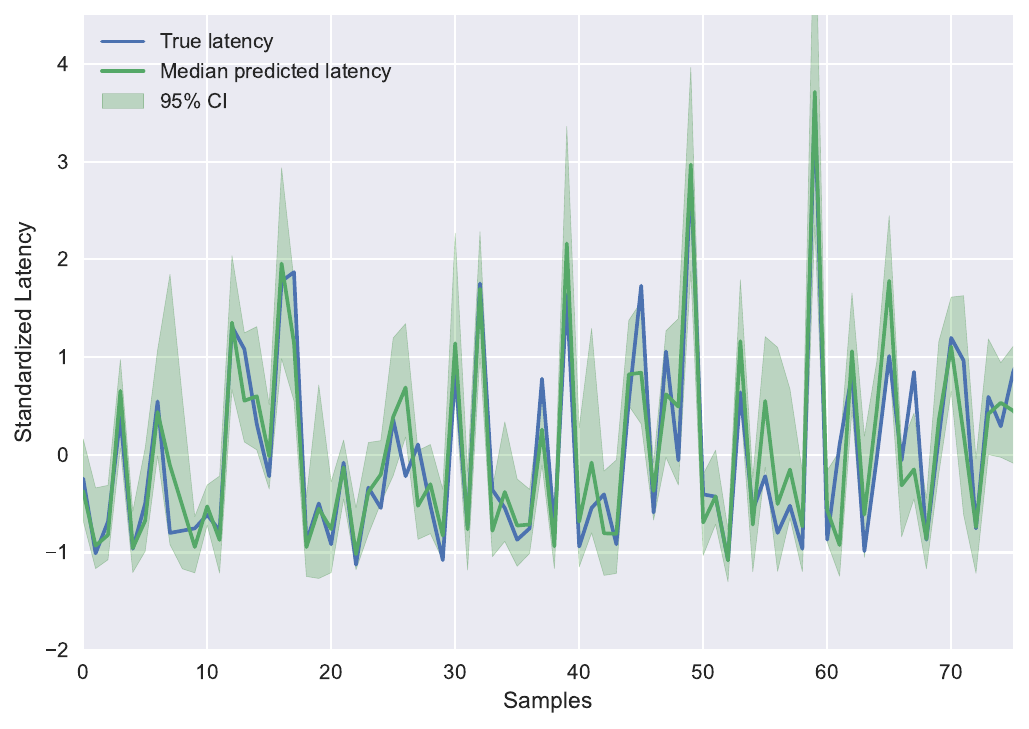}
    \caption{Bayesian Probabilistic Neural Network}
    \label{fig:BPNN}
    \vspace{-.5cm}
\end{figure}

Accurate evaluation of network performance in mobile networks requires the application of regression techniques to \gls{qos} indicators. For instance, these can be employed by \glspl{mno} to assess network performance using simulated data prior to on-field deployment. Unlike non-probabilistic regression methods that provide only a single-point estimate of the predicted value, probabilistic regression allows for the estimation of the probability distribution of the predicted values, providing a complete picture of the underlying uncertainty associated with the predictions. \glspl{bnn} \cite{blundell2015weight}, in particular, are a powerful tool for modeling aleatoric and epistemic uncertainty in a principled way. While the former refers to the intrinsic randomness of the observed data, the latter is captured by the posterior distribution $P(\theta \vert D)$ of the \gls{bnn}'s parametrized model weights, which is updated by means of Bayesian inference as new data $D$ becomes available. In practice, \glspl{bnn} are usually trained via \gls{svi} by minimizing a Monte-Carlo estimate of the variational free energy cost function \eqref{eq:variational_free_energy}\cite{blundell2015weight}:
\begin{equation}\label{eq:variational_free_energy}
\small
    \arg\min_\lambda \left\{KL[q_\lambda(\theta)\Vert P(\theta)]- \mathbb{E}_{\theta \sim q_\lambda}\Big[\log(P(\mathbf{y}\vert \mathbf{x}, \theta))\Big]\right\}.
\end{equation}
In \eqref{eq:variational_free_energy}, the left-hand side term refers to the KL divergence between $q_\lambda$, a variational distribution parametrized by a set of parameters $\lambda$ (typically modeled as a multi-variate normal with learnable diagonal covariance matrix), and $P(\theta)$, the true prior distribution of the model weights. On the right-hand side, $\mathbb{E}_{\theta \sim q_\lambda}[\log(P(\mathbf{y}\vert \mathbf{x}, \theta))]$ refers to the statistical average of the model likelihood, obtained via Monte Carlo sampling of the \gls{bnn}. Minimizing \eqref{eq:variational_free_energy} embodies the tradeoff between maximizing the likelihood over the training data and minimizing the KL divergence with respect to a known prior, which acts as a regularization term. In our experiments, we aim to reflect the latency probability distribution derived in section \ref{latency formulation}. To this end, we explicitly model the last layer of a \gls{bnn} as a Hypoexponential distribution that is parametrized based on the output of the preceding layer. Specifically, the output dimension of the previous layer reflects an $n$-th order approximation of the Hypoexponential distribution. In Fig. \ref{fig:BPNN}, we provide exemplary results on a regression task performed on the dense-urban scenario. As noticeable, the true latency values (blue samples) trustfully lie within the $95\%$ confidence intervals of the probabilistic model, which achieves an overall R2 score of $0.77$ on a held-out test set. It is important to notice that Fig. \ref{fig:BPNN} portrays latency measurements obtained from distinct cells captured at various points in time. Consequently, the depicted data is not arranged in a temporal sequence. 

\subsection{Use case 2: Anomaly detection}
\begin{figure}[t]
    \centering
    \includegraphics[width=\columnwidth]{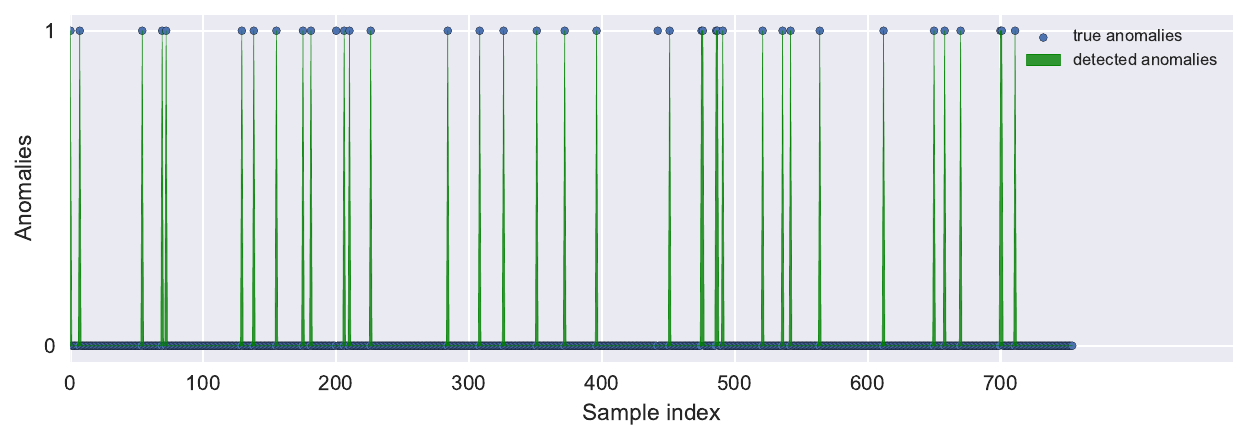}
    \caption{True vs detected anomalies: 36 True positives, 2 False negatives, 0 False positives, 717 True negatives}
    \label{fig:anomaly_detection}
    \vspace{-.7cm}
\end{figure}

Anomaly detection refers to the identification of events significantly differing from the expected behavior of a system. These can manifest by unusual patterns that can be captured or not by network \glspl{kpi}. Typical examples may include network congestion, hardware failures, or jamming attacks.
Based on the assumption that anomalies are often unlikely, the latter can be formulated as a density estimation problem. Given any point $\{\mathbf{x}, \mathbf{y}\} \in \{\mathbb{R}^n, \mathbb{R}^l\}$, if we can estimate a probability density function $\hat{f}_\theta(\mathbf{x},\mathbf{y})$, parametrized by $\theta$, indicating the latency distribution for any given point of the feature space $\in \mathbb{R}^n$, then we can detect an anomaly as per \eqref{eq:anomaly_detection}:
\begin{equation}\label{eq:anomaly_detection}
\small
    \{\mathbf{x}_i, \mathbf{y}_i\} \in \mathcal{A} \Longleftrightarrow \hat{f}(\mathbf{x}_i, \mathbf{y}_i\vert \theta) \leq \Gamma,
\end{equation}
where $\mathcal{A}$ denotes the set of anomalies and $\Gamma$ indicates a likelihood threshold, which is fine-tuned a-posteriori based on a cost model devised as a function of the confusion matrix.\\
When targeting anomaly detection of latency patterns, two distinct methodologies can be pursued, as elaborated upon subsequently: (i) The establishment of a threshold on the \textit{conditional probability distribution} of y given x, i.e., $\hat{f}(\mathbf{x}_i, \mathbf{y}_i\vert \theta) = P(\mathbf{y}\vert\mathbf{x}, \theta)$, which can be suitably modeled using either a \gls{pnn} or a \gls{bnn}, or (ii) The establishment of a threshold based on the reconstruction error of an \gls{ae} on the whole set $\{\mathbf{x},\mathbf{y}\}$. In the latter case, the \textit{joint probability distribution} of x,y, i.e., $\hat{f}(\mathbf{x}_i, \mathbf{y}_i\vert \theta) = P(\mathbf{x}, \mathbf{y}\vert\theta)$ is modeled by the latent space of the \gls{ae}. Both methods are rational as a network's \glspl{kpi} in the feature space $\mathbf{x}$ can detect abnormal situations like network congestion. Conversely, such \glspl{kpi} may not be able to recognize anomalies such as malfunctioning antenna hardware. Empirical findings resulting from the application of the second approach are depicted in Fig. \ref{fig:anomaly_detection}. To construct our test set, we adopt the following method: utilizing the social gathering events dataset, we designate as anomalous the samples obtained from the cellular network coverage encompassing the stadium during the concert event from 7 pm to 11:30 pm on the 16th and 17th of March 2023, yielding a total of 38 anomalies. We subsequently train our \gls{ae} on a set of non-anomalous samples, obtaining a confusion matrix yielding 717 true negatives, 2 false negatives, 0 false positives, and 36 true positives on the test set.

\subsection{Use case 3: Predictive forecasting}
As a final use case, we focus here on predictive forecasting leveraging temporal and spatial information. Predictive forecasting refers to the prediction of future latency given a set of instantaneous and past observations gathered at different locations in the network. This is at the core of \gls{pqos}, as it allows for a proactive optimization approach. In this section, we aim to show how predictive latency forecasting can effectively be achieved by leveraging spatial and temporal information with the use of \glspl{rnn} and \glspl{gnn}. While \gls{lstm} networks \cite{yu2019review} have the ability to capture long-term dependencies in time-series by utilizing a memory cell and three gating mechanisms, \glspl{gnn} afford a strong relational inductive bias beyond that which convolutional and recurrent layers can provide \cite{battaglia2018relational}. In the remainder of the section, we present the numerical outcomes achieved by utilizing a probabilistic \gls{lstm} and GraphSAGE \cite{hamilton2017inductive} on Key Performance Indicators (KPIs) collected in the two distinct scenarios of vehicular mobility and social gathering events.

\subsubsection{Time-series forecasting}
\begin{figure}[h]
    \centering
    \includegraphics[width=.6\columnwidth]{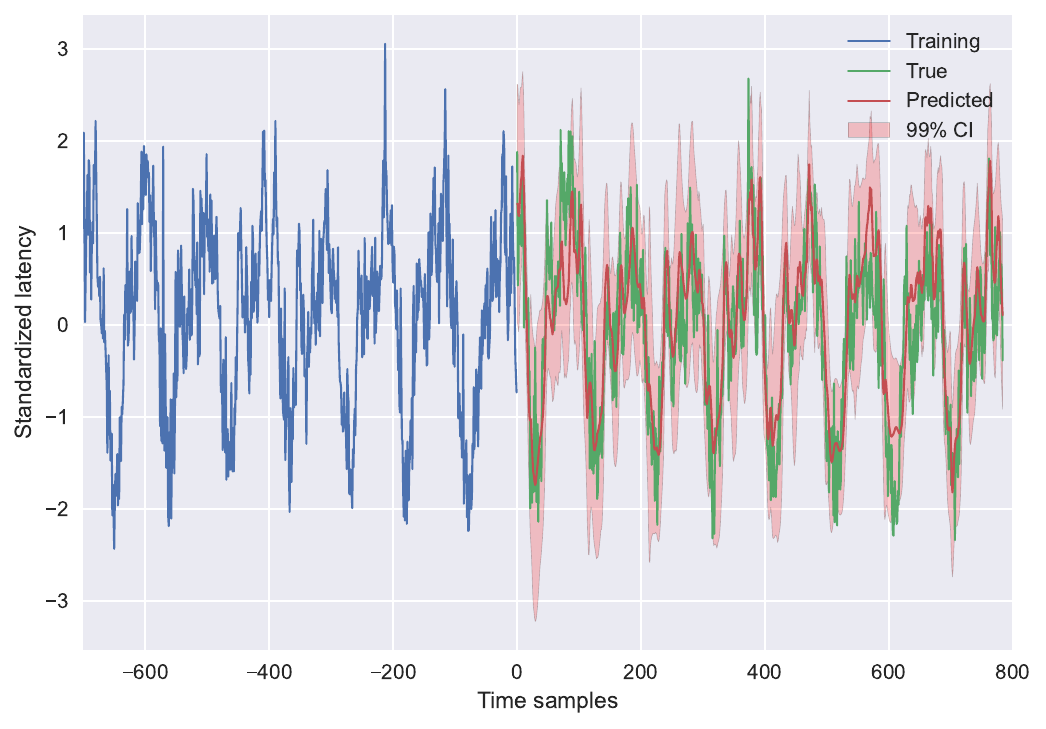}
    \caption{Probabilistic LSTM}
    \label{fig:lstm_results}
    \vspace{-.4cm}
\end{figure}

We leverage a \gls{lstm} model equipped with a probabilistic layer at its final stage and trained via minimization of negative log-likelihood.
For the sake of simplicity, and without loss of generality, we focus on the prediction task of instant $t+1$, i.e. 15 min ahead. The vehicular traffic dataset was partitioned into two sets, with 20 consecutive days designated for training purposes and the subsequent 10 days employed for testing (Fig. \ref{fig:lstm_results}), obtaining an overall R2 score of $0.75$.

\subsubsection{Spatial forecasting}

\begin{figure}[h]
  \centering
  \begin{subfigure}{0.43\columnwidth}
    \includegraphics[width=\linewidth]{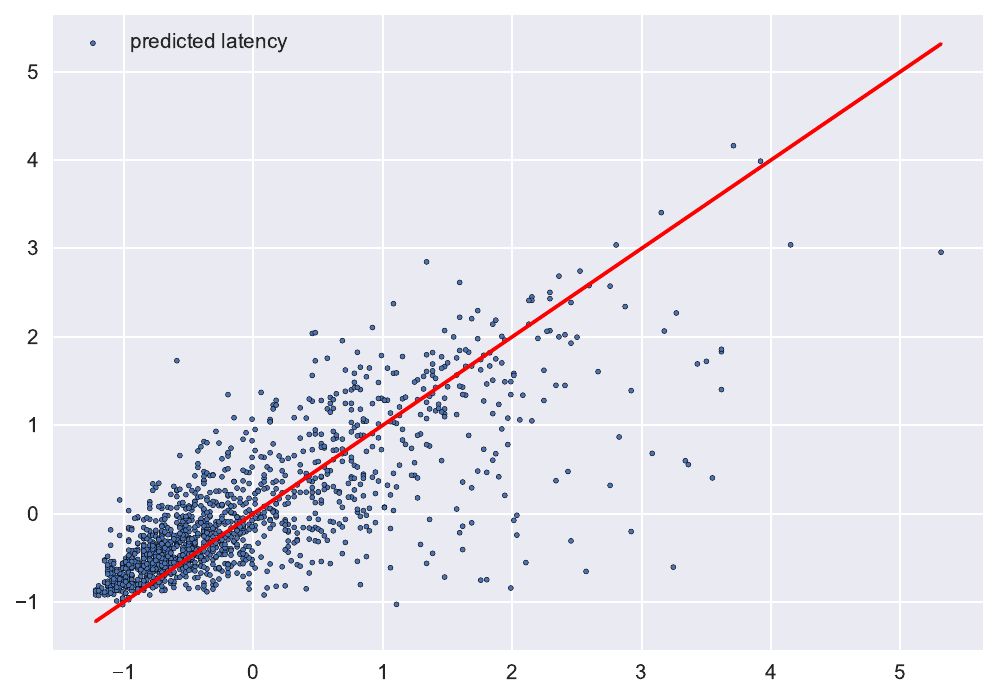}
    \caption{DNN, R2 score = 0.62}
    \label{fig:subfig1}
  \end{subfigure}
  \hspace{0.05\textwidth}
  \begin{subfigure}{0.43\columnwidth}
    \includegraphics[width=\linewidth]{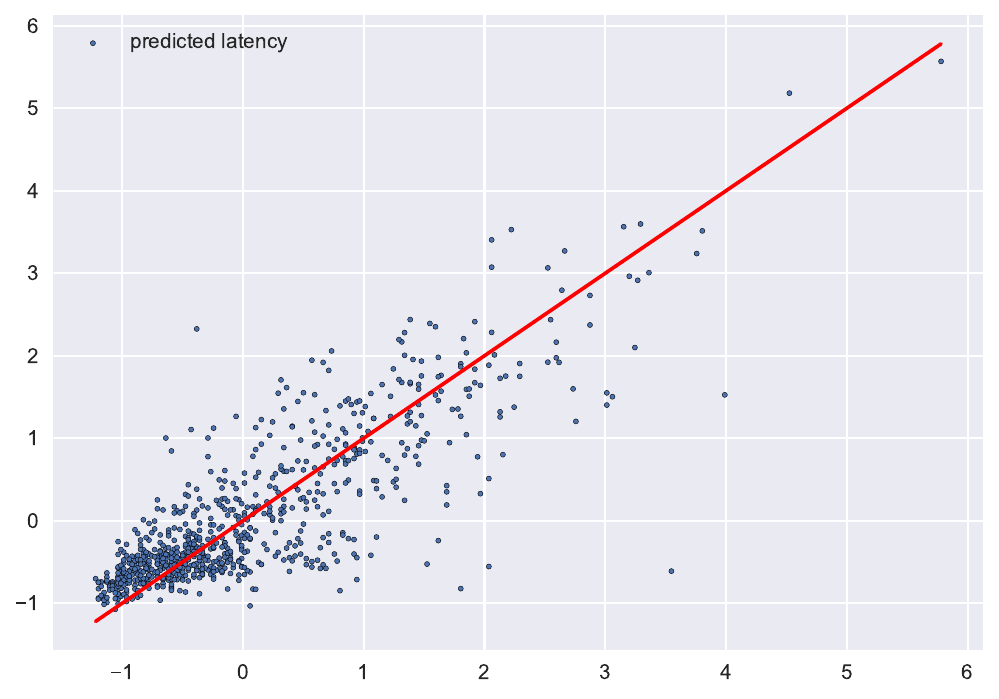}
    \caption{GraphSAGE, R2 score = 0.77}
    \label{fig:subfig2}
  \end{subfigure}
  \caption{GNN vs DNN, predictive latency forecasting}
  \label{fig:subfigure}
  \label{fig:gnn_vs_dnn}
  \vspace{-.4cm}
\end{figure}

Lastly, we compare the performance of a GraphSAGE model, composed of 3 graph convolutional layers, against a baseline \gls{dnn} trained on the entire dataset with samples from individual cells. The obtained results, depicted in Fig. \ref{fig:gnn_vs_dnn}, demonstrate that the former yields superior performance with an R2 score of 0.77 compared to the baseline \gls{dnn} with an R2 score of 0.62. As expected, the obtained results suggest that incorporating spatial information from neighboring data points can significantly enhance the model's predictive capability.

\section{Conclusion}\label{conclusions}

In this work, we provided a comprehensive theoretical and experimental analysis of predictive latency using real-world measurements available to \glspl{mno}. The principal outcomes of our study demonstrate that the latency observed in the U-plane conforms to a Hypoexponential probability distribution. This valuable insight was utilized in our experimental assessment of state-of-the-art \gls{ml} techniques in the context of probabilistic regression, anomaly detection and predictive forecasting.

\appendices
\section{Sojourn time distribution}
The sojourn time $S$ of a packet in a system can be calculated as the sum of its waiting time $W$ in the queue and the service time $B$ required by the server to process the request \cite{kleinrock}. Here, $\tau_{\text{sch}}$ and $\tau_{\text{pack}}$ represent $W$ and $B$, respectively. Under the hypothesis of a $M/M/1$ queue, the packets' arrivals are Poisson distributed with parameter $\beta$, $B\sim exp(\mu)$, and the utilization factor, $\rho :=\beta / \mu \leq 1$.

According to the $Pollaczek \;–Khinchine$ formula for a $M/G/1$ queue \cite{pollak}, the Laplace-Stieltjes transform, $\tilde{S}(s)$ of $S$ is expressed as:

\begin{equation}
\label{laplace}
\small
   \tilde{S}(s) = \frac{(1-\rho) \cdot  \tilde{B}(s) \cdot s}{\beta \cdot  \tilde{B}(s)+s-\beta}.
 \end{equation}
In a $M/M/1$ model, $\tilde{B}(s)=\mu /(\mu+s)$  \cite{kleinrock}. Therefore, Eq. (\ref{laplace}), becomes:

\begin{equation}
\small
\tilde{S}(s) = \frac{\mu \cdot (1-\rho)}{\mu \cdot (1-\rho)+s)}
 \end{equation}
Applying the definition of Laplace-Stieltjes transform of a non-negative r.v $X$, i.e.,$\int_{x=0}^{\infty} e^{-sx} \cdot f(x) dx$ with $s \ge 0$, we obtain:

\begin{equation}
\label{laplace2}
\small
   \int_{x=0}^{\infty} e^{-st} \cdot \lambda e^{-\lambda \cdot t} dt=\frac{\mu \cdot (1-\rho)}{\mu \cdot (1-\rho)+s}= \tilde{S}(s).
 \end{equation}
 Hence, $S$ is exponentially distributed with rate parameter $\lambda = \mu \cdot (1-\rho)$, that is $f_S(t)=\lambda e^{-\lambda \cdot t}$.
\vspace{-.1cm}

\section{Derivation of $L$'s pdf}
For sufficiently small values of $BLER$, and without loss of generality, let us consider the case for which \eqref{eq:P(L2)} can be approximated at the 2-nd order, i.e. $L \approx \tau_{\text{tx}} + \underbrace{\tau_{\text{rtx}} \cdot P_1}_{\tau_{\text{rtx}}'}$. The probability distribution of the sum of two independent continuous random variables can be computed as the convolution between the two individual distributions. Therefore, considering $\tau \sim \tau_{\text{tx}}$ and $\tau_{L} \sim L = \tau_{\text{tx}} + \tau_{\text{rtx}}'$, we have:
\begin{equation*}
\small
\begin{split}
    p_{L_2}(\tau_{L})~=&~\int_{-\infty}^{\infty}p_{\tau_{\text{tx}}}(\tau)p_{\tau_{\text{rtx}}}(\tau_{L} - \tau)~dx~=\\&
    \int_{0}^{\tau_{L}}\lambda_1 e^{-\lambda_1 \tau}\frac{\lambda_2}{P_1} e^{-\frac{\lambda_2}{P_1} (\tau_{L} - \tau)}~dx~=\\&
    \lambda_1 \frac{\lambda_2}{P_1} e^{-\frac{\lambda_2}{P_1} \tau_{L}} \int_{0}^{\tau_{L}} e^{(\frac{\lambda_2}{P_1} - \lambda_1) \tau}~dx~=\\&
    \frac{\lambda_1\lambda_2}{\lambda_2 - \frac{\lambda_1}{P_1}} e^{-\frac{\lambda_2}{P_1} \tau_{L}}\Big[e^(\frac{\lambda_2}{P_1} - \lambda_1)\tau\Big]_0^{\tau_{L}}~=\\&
    \frac{\lambda_1\lambda_2}{\lambda_2 - \frac{\lambda_1}{P_1}}\Big(e^{-\lambda_1 \tau_{L}} - e^{-\frac{\lambda_2}{P_1}\tau_{L}}\Big),
\end{split}
\label{eq:convolution}
\end{equation*}
which is equivalent to the probability distribution of $L \sim hexp(\lambda_1, \lambda_2/P_1)$.\vspace{-.1cm}

\section*{Acknowledgments} 
\small
This work was partially supported by the European Union
under the Italian National Recovery and Resilience Plan
(NRRP) of NextGenerationEU, partnership on “Telecommunications of the Future” (PE00000001 - program “RESTART”).

\bibliographystyle{IEEEtran}
\bibliography{IEEEabrv,StringDefinitions,bibl.bib}

\end{document}